\documentclass[12pt]{article}

\begin{document}

\author{Miguel H. Ib\'{a}\~{n}ez S. \\
Centro de Astrof\'{\i}sica Te\'{o}rica, Facultad de Ciencias,\\
Universidad de los Andes, Apartado de Correos 26,\\
Ipostel, La Hechicera, M\'{e}rida, Venezuela}
\title{Thermochemical Instabilities in Optically Thin Reacting Plasmas}
\maketitle

\begin{abstract}
The linear stability analysis of an optically thin plasma where a general
reaction proceeds, including chemical relaxation time effects, is carried
out . A fifth order dispersion equation (instead of the fourth order one
resulting when such effects are neglected) is obtained. The new mode with
the corresponding instability criterion as well as the modifications of the
previous four modes and the corresponding instability criteria, are
analyzed. Generally, a further stabilizing effect on the unstable modes and
an increasing of the damping of stable modes appear because of the second
viscosity generated by the chemical reaction. The results are applied to:
(1) a collisionally ionized pure hydrogen plasma heated at a constant rate
per unit mass and cooled by free-free transitions, ionization, and $e-H$
collisional excitations; (2) a diffused gas with metallicity $Z$,
photoionized and heated by a radiation field, and cooled by excitation of
hydrogen and heavy metal lines.
\end{abstract}

\section{ Introduction}

It has long been recognized that the knowledge of the thermochemical
equilibrium and stability of plasmas is a crucial point for understanding
the origin of astrophysical structures at very different scales ( Field 1965
). In fact, the onset of thermal ( or more generally thermochemical )
instabilities can directly induce the formation of inhomogeneities or can
set in physical conditions appropriated for developing other kind of
instabilities, say for instance, the gravitational one. As a direct agent in
forming condensations it has been considered in studies of the solar
atmosphere (Priest 1987; Reale, Serio \& Peres 1994 ), the interstellar
medium (Field Goldsmith, Habing \& Field 1969; Hunter 1970; McKee \&
Ostriker 1977; Flannery and Press 1979; Cowie, McKee \& Ostriker 1981;), gas
at high latitude and in outer disks of galaxies ( Verschuur \& Magnani 1994,
Ferrara \& Field 1994 ), Ly$\alpha $ clouds (Sargent et al. 1980; Black
1981; Bond , Szalay \& Silk 1988; Baron et al. 1989; Murakami \& Ikeuchi
1990; Pettini et al. 1990; Duncan, Vishniac \& Ostriker 1991); quasars gas (
Krolik, McKee\& Tarter 1981; Mathews 1986; Mathews and Doane 1990;
Goncalves, Jatenco-Pereira \& Opher 1993 ), and as an indirect promoter of
the formation of stars ( Glassgold and Lager 1976; Oppenheimer 1977; Sabano
and Kannari 1978; Iba\~{n}ez 1981 ), globular clusters ( Fall and Rees 1985;
Murray \& Lin 1990a,b ) and galaxies ( Sunyaev and Zel'dovich 1972; Gurevich
\& Chernin 1975; Zel'dovich \& Novikov 1983 ).

On the other hand, the generalization of Field's (1965 ) basic study on
thermal instability for including effects of chemical reactions has been
carried out by Goldsmith (1970), Defouw (1970), Yoneyama (1973), Flannery
and Press 1979; Iba\~{n}ez \& Parravano (1983); Iba\~{n}ez \& Mendoza
(1990), and Corbelli and Ferrara (1995 ). However, in these works it has not
been taken into account the fact that when the temperature, density and the
other thermodynamical quantities change, the position of the chemical
equilibrium also varies, i.e. in a reacting gas the fluctuations occur on
states out of thermodynamic equilibrium because, generally, the chemical
relaxation time is not short enough to follow the change of pressure and
density. So, if the restoring of the chemical equilibrium after an initial
compression, for instance, occurs relatively slowly, it can not follow the
compression. But the irreversible tendency of the concentrations to reach
the equilibrium values corresponding to the new value of pressure and
density produces an increase of entropy with the corresponding energy
dissipation. ( Landau and Lifshitz 1987). The present work is aimed at
considering such effects in a general reacting gas. Additionally, two
particular applications will be carried out: (1) to the pure hydrogen plasma
model worked out by Iba\~{n}ez \& Parravano (1983), and (2) to the
photoionized plasma with metallicity $Z$ studied by Corbelli and Ferrara
1995 , hereinafter references IP and CF, respectively.

\section{Basic Equations}

For a fluid where a chemical reaction of the form $\sum \nu _iC_i=0$ is
proceeding, the equations of gas dynamics can be written in the form 
\begin{equation}
\frac{\partial \rho }{\partial t}+\mathbf{\nabla \cdot (\rho v})=0~,
\label{1}
\end{equation}
\begin{equation}
\rho \frac{d\mathbf{v}}{dt}+\mathbf{\nabla }p=0~,  \label{2}
\end{equation}
\begin{equation}
\frac{d\xi }{dt}+X(\rho ,T,\xi )=0~,  \label{3}
\end{equation}
\begin{equation}
AR\frac{dT}{dt}-\frac p{\rho ^2}\frac{d\rho }{dt}+BRT\frac{d\xi }{dt}+%
\mathcal{L}(\rho ,T,\xi )-\frac 1\rho \mathbf{\nabla } \bf{\cdot (}\kappa 
\mathbf{\nabla }T)=0~,  \label{4}
\end{equation}
\begin{equation}
p(\rho ,T,\xi )=R\frac{\rho T}{\mu (\xi )}~,  \label{5}
\end{equation}
where $\frac d{dt}$ is the convective derivative and the remaining symbols
are defined in Table 1.

For reacting fluids is useful to rewrite equation (3) in the form 
\begin{equation}
\frac{d\xi }{dt}=\frac{\phi (\rho ,T,\xi )}{\tau (\rho ,T,\xi )}~,  \label{6}
\end{equation}
where 
\begin{equation}
\tau (\rho ,T,\xi )=\left[ (\frac{\partial X}{\partial \xi })_{\rho
,T}\right] ^{-1}=\frac 1{X_\xi }~,  \label{7}
\end{equation}
and 
\begin{equation}
\phi (\rho ,T,\xi )=\frac{X(\rho ,T,\xi )}{(\frac{\partial X}{\partial \xi }%
)_{\rho ,T}}~,  \label{8}
\end{equation}
see for instance Landau and Lifshitz (1987), Zel'dovich and Raizer (1966),
Vicenti and Kruger (1975). When the reacting fluid is chemically stable $%
X_\xi >0,$ $\tau $ becomes the chemical relaxation time, otherwise, $\tau $
is the $e-folding$ time for chemical instability.

The chemical equilibrium is defined by the condition 
\begin{equation}
\phi (\rho ,T,\xi )=0~,  \label{9}
\end{equation}
therefore, the chemical equilibrium value of $\xi $ ( denoted by $\xi _{*%
 }$) becomes a function of the density and the temperature which
generally are functions of time, i.e. $\xi _{*}=\xi _{*}[\rho (t),T(t)]$.

On the other hand, and as it is well known, in an hypothetical infinite
medium, equations (1)-(5) admit a steady state (uniform ) solution
corresponding to a complete thermodynamical equilibrium, i.e. $\mathbf{v}=0,$
and $\rho $ and $T$ satisfying the equation

\begin{equation}
\mathcal{L}(\rho ,T,\xi )=0~.  \label{10}
\end{equation}
Equation (10) defines the thermal equilibrium.

The stability of the above equilibria is usually analyzed ( in the linear
approximation ) looking for solutions of equations (1)-(5) in the form of
plane waves, superimposed to the equilibrium values i.e.

\begin{equation}
\delta \psi =\psi _0+\psi ^{\prime }(t)\exp (i\mathbf{k\cdot r})~,
\label{11}
\end{equation}
where $\psi _0$ is the $\psi $ value at equilibrium and $\psi ^{\prime }(t)$
is the respective time dependence of the disturbance on the variable $\psi $%
. Substituting equation (11) in equations (1)-(5) and neglecting nonlinear
terms in $\psi ^{\prime }$ one obtains the set of linearized equations 
\begin{equation}
\frac{\partial \rho ^{\prime }}{\partial t}+i\rho _0\mathbf{k\cdot v}%
^{\prime }=0,  \label{12}
\end{equation}
\begin{equation}
\rho _0\frac{\partial \mathbf{v}^{\prime }}{\partial t}+i\mathbf{k}p^{\prime
}=0~,  \label{13}
\end{equation}
\begin{equation}
\tau \frac{\partial \xi ^{\prime }}{\partial t}+\xi ^{\prime }=\xi _{*\rho
}\rho ^{\prime }+\xi _{*T}T^{\prime }~,  \label{14}
\end{equation}
\begin{equation}
-\frac{p_0}{\rho _0^2}\frac{\partial \rho ^{\prime }}{\partial t}+AR\frac{%
\partial T^{\prime }}{\partial t}+BRT_0\frac{\partial \xi ^{\prime }}{%
\partial t}+\mathcal{L}_\rho \rho ^{\prime }+\mathcal{L}_TT^{\prime }+%
\mathcal{L}_\xi \xi ^{\prime }+\frac{\kappa _0k^2}{\rho _0}T^{\prime }=0~,
\label{15}
\end{equation}
\begin{equation}
p^{\prime }=p_\rho \rho ^{\prime }+p_TT^{\prime }+p_\xi \xi ^{\prime }~.
\label{16}
\end{equation}
Hereinafter, the subindex $_0$ denoting equilibrium values will be dropped.

The linearized equation (14) takes into account the fact that when $\rho $
and $T$ change, the position of the equilibrium $\xi _{*}$ also varies (
Landau and Lifshitz 1987; Zel'dovich and Raizer 1966 ). For a detailed
analysis of this basic aspect of the problem, see Vincenti and Kruger (1975).

As usually, one may proceed further looking for solutions of the dynamical
variables $\mathbf{v}^{\prime }(t)$, $\rho ^{\prime }(t)$ and $T^{\prime
}(t)\sim \exp (\mathcal{N}t).$ However, from equation (14) follows that a
solution for $\xi ^{\prime }\sim \exp (\mathcal{N}t)$ exists if and only if
the amplitude of the disturbances $\xi _1$, $\rho _{1 }$and $T_{1%
 }$are related by the equation 
\begin{equation}
\xi _1=\frac{\xi _{*\rho }}{\mathcal{N}\tau +1}\rho _1+\frac{\xi _{*T}}{%
\mathcal{N}\tau +1}T_1~,  \label{17}
\end{equation}
where $\rho _1$ and $T_1$ are the amplitude of the density and temperature
fluctuation, respectively. Therefore, equations (12)-(16) are reduced to two
algebraic equations for $\rho _{1 }$ and $T_1,$ i. e.

\begin{equation}
\lbrack (\mathcal{N}\tau +1)(\mathcal{N}^2+p_\rho k^2)+\xi _{*\rho }p_\xi
k^2)]\rho _1+[(\mathcal{N}\tau +1)p_Tk^2+\xi _{*T}p_\xi k^2]T_1=0~,
\label{18}
\end{equation}
\[
\lbrack (\mathcal{N}\tau +1)(-\frac p{\rho ^2}\mathcal{N}+\mathcal{L}_\rho
)+\xi _{*\rho }(BRT\mathcal{N}+\mathcal{L}_\xi )]\rho _1+[(\mathcal{N}\tau
+1)(AR\mathcal{N}+\mathcal{L}_T+\frac{\kappa k^2}\rho )+. 
\]

\begin{equation}
\xi _{*T}(BRT\mathcal{N}+\mathcal{L}_\xi )]T_1=0,  \label{19}
\end{equation}
which have non trivial solution provided that the compatibility condition 
\begin{equation}
a_0\emph{N}^5+a_1\emph{N}^4+a_2\emph{N}^3+a_3\emph{N}^2+a_4\emph{N}+a_5=0~,
\label{20}
\end{equation}
holds, where

\[
a_0=\tau ^2~,\quad a_1=\tau (2+\frac BAT\xi _{*T})+\tau ^2ck_1~, 
\]
\[
a_2=(1+\frac BA\xi _{*T}T)+\tau c(2k_1+k_3)+\tau ^2c^2k^2~, 
\]
\begin{equation}
a_3=c(k_1+k_3)+\tau \frac{\Gamma _1c^2}{\tilde{\gamma}}k^2+\tau ^2\frac{c^3}{%
\tilde{\gamma}}(k_1-k_2)k^2~,  \label{21}
\end{equation}
\[
a_4=\frac{\Gamma _2c^2}{\tilde{\gamma}}k^2+\tau \frac{c^3}{\tilde{\gamma}}[%
(2+\xi _{*\rho }\rho \mu \nu )k_1-(2+\xi _{*T}T\mu \nu )k_2+(1-\frac{\rho
\xi _{*\rho }}{T\xi _{*T}})k_3]k^2~, 
\]
\[
a_5=\frac{c^3}{\tilde{\gamma}}[(1+\xi _{*\rho }\rho \mu \nu )k_1-(1+\xi
_{*T}T\mu \nu )k_2+(1-\frac{\rho \xi _{*\rho }}{T\xi _{*T}})k_3]k^2~. 
\]

In previous works ( Goldsmith 1970; Defouw 1970; Yoneyama 1970; Flannery and
Press 1979, IP and CF ) where the change in the chemical equilibrium
position was not taken into account, the characteristic equation found was a
fourth order , instead of a fifth order polynomial, equation (20). Note that
the way of considering the disturbance in the chemical parameter $\xi $ in
eqs.(14) and (17) is different from that of the above papers. In fact, in
those papers the hypothetical initial ''equilibrium state'' is considered 
\emph{frozen}, as that in absence of relaxing processes and the chemical
fluctuation is assumed to follow that of pressure,density and velocity.
However, this is only an acceptable first approximation. In deed, once the
initial equilibrium is disturbed, an irreversible process set in and the
chemical evolution tends to a \emph{new equilibrium} corresponding to the 
\emph{new values} of pressure and density. Therefore, it is not matter how
tiny this effect is, it changes qualitatively the problem. The effect of
finite value of the relaxation time is equivalent to think on a\emph{\
second viscosity }(Landau \& Lifshitz 1987) and this is the last physical
reason by which fluctuations with large wave numbers can be damped in a
fluid where no other irreversible process, except the chemical reaction,
goes into play. This aspect of the problem will be again considered later on.

\section{Instability Criteria}

\subsection{Asymptotic Cases}

Equation (17) gives the correct amplitude for the chemical parameter in both
asymptotic limits: no reacting gases, i.e. when $\tau \rightarrow \infty $, $%
\xi _1\rightarrow 0$, and in \emph{equilibrium flow} for which $\tau
\rightarrow 0$ and $\xi _1\rightarrow \xi _{*\rho }\rho _1+\xi _{*T}T_1$,
i.e. when the chemical reaction is so fast that it instantaneously adjusts
to the chemical equilibrium values and therefore, the changes of density and
temperature occur in chemical equilibrium. Note that in the equilibrium flow
limit, $\phi =0$, $\tau =0$ but $(d\xi /dt)_{eq.}\neq 0$, contrary to the
simple \emph{equilibrium }( as in a closed system with fixed conditions) for
which $\phi =0$ and $(d\xi /dt)_{eq.}=0$, case considered in the mentioned
references at the end of the previous Section. In practice however, and from
the physical point of view, the above asymptotic limits correspond ,
respectively, to the case when the chemical time scale is much longer, ( $%
\mid $\emph{N}$\tau \mid >>1$) and much more shorter ( $\mid $\emph{N}$\tau
\mid <<1$ ) , than the characteristic time of the density and temperature
fluctuation.

On the other hand, as it can be readily verified, when $\tau \rightarrow
\infty $, the dispersion relation (20) reduces to the corresponding Field
(1965) equation, and when $\tau \rightarrow 0$, equation (20) reduces to 
\begin{equation}
a_2\mathcal{N}^3+a_3\mathcal{N}^2+a_4\mathcal{N}+a_5=0~,  \label{22}
\end{equation}
where $a_2$ 
\begin{equation}
a_2=1+\frac BAT\xi _{*T}~;\quad a_3=c(k_1+k_3)~;\quad a_4=\frac{\Gamma _2c^2%
}{\tilde{\gamma}}k^2~,  \label{23}
\end{equation}
and $a_5$ stands as in equation (21).

By applying the Hurwitz criteria to the equation (22) one obtains that an
instability sets in when any of the following relations is fulfilled, 
\begin{equation}
\mathcal{L}_T+\xi _{*T}\mathcal{L}_\xi +\frac{\kappa k^2}\rho \leq 0~,
\label{24}
\end{equation}
\begin{equation}
(1+T\mu \nu \xi _{*T})[(A+BT\xi _{*T})\mathcal{L}\rho +\frac T{\mu \rho
}(1-B\rho \mu \xi _{*\rho })\mathcal{L}_T+(A\xi _{*\rho }+\frac T{\mu \rho
}\xi _{*T})\mathcal{L}_\xi +\frac T\mu (1-B\rho \mu \xi _{*\rho })\frac{%
\kappa k^2}{\rho ^2}]\leq 0~,  \label{25}
\end{equation}
\begin{equation}
(1+\xi _{*\rho }\rho \mu \nu )(T\mathcal{L}_T+T\xi _{*T}\mathcal{L}_\xi
)-(1+\xi _{*T}T\mu \nu )(\rho \mathcal{L}_\rho +\rho \xi _{*\rho }\mathcal{L}%
_\xi )+(1+\xi _{*\rho }\rho \mu \nu )\frac{T\kappa k^2}\rho \leq 0~.
\label{26}
\end{equation}
Note that in absence of thermal conduction the above criteria become
independent of the wave number $k$.

Criterion (24) is Parker's ( 1953) isochoric criterion of thermal
instability but modified by the presence of the chemical reaction, i.e. the
change with time of temperature occurs in chemical equilibrium.

Criterion (25) can be identified as Field's (1965) isentropic criterion for
thermal instability also modified by the chemical reaction. In fact, as can
be readily verified, if the equilibrium value $\xi _{*}(\rho ,T)$ is weekly
dependent on density and temperature, the relation (25) reduces to the
corresponding Field's isentropic criterion. In the opposite case when the
dependence of the heating/cooling function on $\rho $ and $T$ is very week ( 
$\mathcal{L}_\rho \sim \mathcal{L}_T\sim 0$ ), the criterion stands barely
due to internal processes ( the chemical reaction\ ), i.e. 
\begin{equation}
(1+T\mu \nu \xi _{*T})[(A\xi _{*\rho }+\frac T{\mu \rho }\xi _{*T})\mathcal{L%
}_\xi +\frac T\mu (1-B\rho \mu \xi _{*\rho })\frac{\kappa k^2}{\rho ^2}]\leq
0~.  \label{27}
\end{equation}
If $(1+T\mu \nu \xi _{*T})>0$, the criterion simplifies further, that is,
isentropic instability sets in when the expression in square brackets in
relation (27) is less that zero. If the above occurs when $(A\xi _{*\rho
}+\frac T{\mu \rho }\xi _{*T})\mathcal{L}_\xi <0$ and $(1-B\rho \mu \xi
_{*\rho })>0$, sound waves with wave number smaller than the critical number 
$k_{c }$defined by the equality in (27) are amplified and those
larger ones are damped, but if $(A\xi _{*\rho }+\frac T{\mu \rho }\xi _{*T})%
\mathcal{L}_\xi <0$ and $(1-B\rho \mu \xi _{*\rho })<0,$ sound waves are
amplified regardless of the wave length. On the contrary, if the expression
in square brackets in (27) is positive, the sounds waves are damped.

Relations (26) can be identified as Field's (1965) isobaric criteria for
thermal instability, modified by the chemical reaction. As in the previous
case, the respective thermal instability would proceed in chemical
equilibrium. Generally, from the above modified criterion it follows that
very fast chemical reactions may help to develop or may quench the
respective isobaric thermal instability depending on the signs of $\xi _{*T}%
\mathcal{L}_\xi $ and $\xi _{*\rho }\mathcal{L}_\xi $ terms for the
particular reaction under consideration. In particular, if $(1/\mu +\xi
_{*\rho }\rho \nu )>0$, this criterion can be written as 
\begin{equation}
(T\mathcal{L}_T-\beta \rho \mathcal{L}_\rho )+(T\xi _{*T}-\beta \rho \xi
_{*\rho })\mathcal{L}_\xi +\frac{T\kappa k^2}\rho \leq 0~,  \label{28}
\end{equation}
where 
\begin{equation}
\beta =\frac{1/\mu +\xi _{*T}\rho \nu }{1/\mu +\xi _{*\rho }\rho \nu }~.
\label{29}
\end{equation}
Note that if the chemical parameter in chemical equilibrium $\xi _{*}$
weekly depends on density and temperature, i.e. $\xi _{*T}\sim \xi _{*\rho
}\sim 0$, the above criterion reduces to the isobaric Field's criterion for
nonreacting gases. However, the above generalized criterion, relation (28),
provides further information. In effect, in the opposite case, that is, when
the heating/cooling function $\mathcal{L}$ weekly depends on $\rho $ and $T$%
, $\mathcal{L}_T\sim \mathcal{L}_\rho \sim 0$ , criterion (28) simplifies to 
\begin{equation}
(T\xi _{*T}-\beta \rho \xi _{*\rho })\mathcal{L}_\xi +\frac{T\kappa k^2}\rho
\leq 0~,  \label{30}
\end{equation}
which can be identified as an isobaric instability criterion due only to the
chemical reaction .

Other important conclusion drawn from the instability criteria (24)-(26) is
concerned with the critical wave numbers $k_c$ defined by the respective
equality which corresponds to the marginal states. It is obvious that very
fast chemical reactions may produce drastic changes in the critical scale
lengths at which any thermal instability may occur. In particular, if $%
\mathcal{L}_T<0$, and $\xi _{*T}\mathcal{L}_\xi >0$ but this last term is
not large enough for quenching the isochoric instability, the critical wave
number $\lambda _c$ ( $=2\pi /k_c$ )can be considerable increased respect to
the critical wave number in a non-reacting gas. Similar considerations can
be made for the remaining two critical wave numbers, the isentropic and the
isobaric ones, as it will be seen bellow where particular applications will
be considered.

On the other hand, for fluctuations with wave number $k=0$, equation (20)
becomes 
\begin{equation}
\emph{N}^2(a_0\emph{N}^3+a_1\emph{N}^2+a_2\emph{N}+a_3)=0~,  \label{31}
\end{equation}
where $a_0$ and $a_{1 }$stand as in equation (21), but $a_2$ and $a_3$
reduce to 
\begin{equation}
a_2=(1+\frac BA\xi _{*T}T)+\tau c(2k_1^{\prime }+k_3)~;\quad
a_3=c(k_1^{\prime }+k_3)~.  \label{32}
\end{equation}
If in addition $\tau =0$ , the dispersion relation (31 ) further reduces to 
\begin{equation}
\emph{N}^2(a_2\emph{N}+a_3)=0~,  \label{33}
\end{equation}
i.e. there is only one ( thermoreactive ) mode with rate $\emph{N}%
=-c(k_1^{\prime }+k_3)/(1+B\xi _{*T}T/A)$, which becomes unstable if $%
k_1^{\prime }+k_3<0$, provided that $1+B\xi _{*T}T/A>0$.

For the opposite asymptotic case, i.e. when $k\rightarrow \infty $, the
dispersion equation (20) simplifies to 
\begin{equation}
\mathcal{N}^2+\frac 1\tau (2+\xi _{*\rho }\rho \mu \nu )\mathcal{N}+\frac
1{\tau ^2}(1+\xi _{*\rho }\rho \mu \nu )=0~.  \label{34}
\end{equation}
Therefore, there are two modes with rates $\mathcal{N}=-1/\tau $ and $%
\mathcal{N}=-(1+\xi _{*\rho }\rho \mu \nu )/\tau $, respectively. Obviously,
for chemically stable reactions the last mode can be unstable provided that $%
(1+\xi _{*\rho }\rho \mu \nu )<0$, otherwise any fluctuation with very short
wave length is damped in a chemical time scale.

\subsection{General Case}

In the general case, when the time scale $\tau $ is $\neq 0$, case usually
found in practical problems, applying the Hurwitz criteria to the fifth
order polynomial (20), five instability criteria are found. Explicitly,
first instability criterion, $a_1\leq 0$, ( for $\tau \neq 0$ ) becomes 
\begin{equation}
\lbrack \mathcal{L}_T\pm \frac{R(2A+\xi _{*T}BT)}{\mid \tau \mid }]+\frac{%
\kappa k^2}\rho \leq 0~,  \label{35}
\end{equation}
with the sign $+$ for chemical stable and the sign $-$ for chemically
unstable reacting fluids. The above criterion reduces to the isochoric
criterion for nonreacting fluids when $\mid \tau \mid \rightarrow \infty $.
On the other hand, for fluctuations with $k=0$, the instability criterion
reduces to the condition $[\mathcal{L}_T\pm R(2A+\xi _{*T}BT)/\mid \tau \mid
]\leq 0$, i.e. for a thermally unstable fluid ( $\mathcal{L}_T<0$ ) the
stabilizing effect of the chemical reaction increases when the chemical time
scale $\mid \tau \mid $ decreases, provided that $2A+B\xi _{*T}T>0$.

Fifth instability criterion is just $a_5\leq 0$, and coincides with the
isobaric modified criterion (28). Therefore, the above criterion is
independent of the chemical time scale $\tau $ and it stands as far as $\tau
\neq \infty $, i.e. when chemical reactions proceed in the fluid under
consideration.

Second, third and fourth instability criteria become, respectively,

\begin{equation}
\Delta _2=a_1a_2-a_0a_3\leq 0~,  \label{36}
\end{equation}
\begin{equation}
\Delta _3=\Delta _2a_3+(a_0a_5-a_1a_4)a_1\leq 0~,  \label{37}
\end{equation}
\begin{equation}
\Delta _4=\Delta _3a_4+[-\Delta _2a_2+a_0(a_1a_4-a_0a_5)]a_5\leq 0~.
\label{38}
\end{equation}
The relation (36) can be explicitly written as 
\[
\frac 1\tau \{(2+\frac BAT\xi _{*T})[c^2k^2+\frac 1{\tau ^2}(1+\frac BAT\xi
_{*T})]-\frac{\Gamma _1c^2k^2}{\tilde{\gamma}}\} 
\]
\[
+\{c^3(1-\frac 1{\tilde{\gamma}})k^2+\frac c{\tau ^2}[\frac BAT\xi
_{*T}+2(2+\frac BAT\xi _{*T})]\}k_1 
\]
\begin{equation}
+\frac 2\tau c^2k_1^2+\frac{c^2}\tau k_1k_3+\frac{c^3k^2}{\tilde{\gamma}}%
k_2+\frac c{\tau ^2}(1+\frac BAT\xi _{*T})k_3\leq 0~,  \label{39}
\end{equation}
which reduces to the isentropic Field's criterion when $\tau \rightarrow
\infty $.\ Therefore, it can be identified as a generalized isentropic
criterion for thermochemical instability. For fluctuations with $k=0$, but
values of $\tau \neq 0$, the above criterion simplifies to

\begin{equation}
\frac 1\tau (2+\frac BAT\xi _{*T})(1+\frac BAT\xi _{*T})+(4+3\frac BAT\xi
_{*T}+2\tau ck_1)ck_1+(1+\frac BAT\xi _{*T}+\tau ck_1)ck_3\leq 0~,
\label{40}
\end{equation}
which in the asymptotic case when $\mathcal{L}_T\approx 0$ becomes 
\begin{equation}
(1+\frac BAT\xi _{*T})[ck_3+\frac 1\tau (2+\frac BAT\xi _{*T})]\leq 0~,
\label{41}
\end{equation}
or simply 
\begin{equation}
\xi _{*T}\mathcal{L}_\xi \pm \frac R{\mid \tau \mid }(2A+BT\xi _{*T})\leq 0~,
\label{42}
\end{equation}
if $(1+BT\xi _{*T}/A)>0$, where $+$ and $-$ signs corresponds to chemically
stable and instable reactions, respectively.

Explicit relations for the criteria (37) and (38) are rather involved and
generally, one must proceed numerically in specific applications. However,
in the asymptotic case for $k=0$, criterion (37) reduces to

\[
\left( 2\,k_1^2+3\,k_1k_3+\,k_3^2\right) c^3k_1\tau ^3+ 
\]

\[
\left( 5\,k_1\,k_3+\frac{4BT\xi _{*T}\,k_1\,k_3}A+k_3^2+4\,k_1^2+\frac{BT\xi
_{*T}\,k_3^2}A+\frac{3BT\xi _{*T}k_1^2}A\right) c^2\tau ^2 
\]

\begin{equation}
\left( 2\,k_1+\frac{3BT\xi _{*T}\,k_1}A+\frac{3BT\xi _{*T}\,k_3}A+2\,k_3+%
\frac{B^2T^2\xi _{*T}^2k_1}{A^2}+\frac{B^2T^2\xi _{*T}^2k_3}{A^2}\right)
c\tau \leq 0\quad .  \label{43}
\end{equation}
If additionally $k_1\approx 0$, the criterion simplifies to

\begin{equation}
c\tau k_3[(1+\frac{BT\xi _{*T}}A)c\tau k_3+(2+\frac{3BT\xi _{*T}}A+\frac{%
B^2T^2\xi _{*T}^2}{A^2})]\leq 0  \label{44}
\end{equation}
which can be identified as a thermoreactive criterion for instability.

Fourth criterion (38) can also be identified as a second thermoreactive
criterion for instability, as it can be verified considering the asymptotic
case $\mathcal{L}_\rho \approx \mathcal{L}_T\approx 0$, but $\mathcal{L}_\xi
\neq 0$.

\section{Astrophysical Applications}

\subsection{Collisionally Ionized Hydrogen Plasma}

This section will be devoted to applying the general results obtained in
sections 2 and 3 to the hydrogen plasma model studied in a previous work
(IP), i.e. to a collisionally ionized pure hydrogen plasma with a net rate
function $X(\rho ,T,\xi )$ and a net cooling rate per unit mass $\mathcal{L}%
(\rho ,T,\xi )$, respectively, given by

\begin{equation}
X(\rho ,T,\xi )=N_0\alpha _B(T)\rho \xi ^2-N_0q(T)\rho \xi (1-\xi )\quad ,
\label{45}
\end{equation}
\begin{equation}
\mathcal{L}(\rho ,T,\xi )=N_0R\rho \xi ^2T\beta _B(T)+N_0^2\chi \rho \xi
(1-\xi )[q(T)+\Phi (T)]\emph{-}\mathcal{L}_0\quad ,  \label{46}
\end{equation}
where $N_0$ is the Avogadro Number, $\chi =13.598$ $e.v$, $\mathcal{L}_0$ is
a constant per unit mass heating and the coefficients $\alpha _B(T)$, $q(T)$%
, $\beta _B(T)$ and $\Phi (T)$ are given by Seaton (1959), Hummer and Seaton
(1963) and Hummer (1963). The galactic value $\mathcal{L}_0=3.25\times
10^{-4}$ $ergs$ $g^{-1}$ $s^{-1}$ ( Potasch, Wesselius and Duinen 1979 ) has
been taken as a reference value. Additionally, the thermal conduction
coefficient is taken in the form 
\begin{equation}
\kappa (\rho ,T,\xi )=2.5\times 10^3(1-\xi )T^{1/2}+1.84\times 10^{-5}\frac{%
\xi T^{5/2}}{\ln \Lambda (\rho ,T,\xi )}\quad ,  \label{47}
\end{equation}
where the first term on the right hand side is the heat conduction by
neutral atoms ( Parker 1953) and the second one is the Spitzer ( 1962 )
thermal conduction by electrons, where

\begin{eqnarray}
\ln \Lambda &=&23.24+\ln [(\frac T{10^4})^3(\frac 1{N_0\rho \xi })^{1/2}]%
, if T<4.2\times 10^5K,  \label{48} \\
\ln \Lambda &=&29.71+\ln [\frac T{10^6(N_0\rho \xi )^{1/2}}] , \quad ~
~if T>4.2\times 10^5K.  \nonumber
\end{eqnarray}
The range of temperature under consideration is $3\times 10^3<T<8\times
10^5K $ .

The equilibrium pressure as a function of particle number density has been
plotted in Figure 1, on which the value of temperature of the marginal
states have been also indicated. From solving the IP dispersion equation
follows that regardless the value of $k$ the hydrogen plasma is
thermoreactivelly unstable in the range of temperature $T_3$ ( $=1.66\times
10^4$ $K$ )$<T<T_4$ ( $=4.30\times 10^5$ $K$ ), with a maximum growing rate
of $3.2\times 10^{-9}$ $yr^{-1}$ at a temperature $T=3.2\times 10^4$ $K$ .
Additionally, there is an oscillatory mode which, for wave numbers $k\leq
2.1\times 10^{-4}$, is isobarically unstable in two temperature ranges ,
close to $T\sim 9\times 10^3$ $K$ and to $T\sim 5\times 10^4$ $K$ , see Fig
2a. The lower range of temperature increases when $k$ decreases. In Fig. 1
such a range $T_1$ ( $=8.65\times 10^3$ $K$ )$<T<T_2$ ( $=1.26\times 10^4$ $%
K $ ) is shown for the asymptotic value $k\rightarrow 0.$ For $k\geq
1.7\times 10^{-3}$ this oscillatory mode stabilizes$.$ The corresponding
maximum growing rate is very short. For instance, a disturbance with $%
k=10^{-4}$, reaches a maximum growing rate ${Re}\{n_{\max }\}$ $%
=7.8\times 10^{-12}$ $yr^{-1}$ at $T=9.66\times 10^3$ $K$ at the lower
temperature unstable range, and ${Re}\{n_{\max }\}$ $=4.4\times
10^{-11} $ $yr^{-1}$ at $T=3.77\times 10^4$ $K$, for the parameter values
used in Fig.1. Note that $k$ is normalized to the value $k_{*}=1.89\times
10^5\mathcal{L}_0/RT_1c_{*}$, where $c_{*}=(\frac 53RT_1)^{1/2}$ and $%
T_1=157890K $. Therefore, for the galactic value taken by IP as a reference
value, $\mathcal{L}_0=3.25\times 10^{-4}$ $erg$ $g^{-1}$ $s^{-1}$, one gets $%
k_{*}=10^{-18}$ $cm^{-1}$. For disturbances with larger wave numbers the
oscillatory modes are damped.

Figure 2b is a plot of the resulting rates from solving the dispersion
equation (20). Qualitatively, the above results obtained by IP hold.
However, the maximum values of the growing rates increase by a factor $2$
and the fastest real damped mode ( for $T>10^4$ $K$ , in Fig 2a ) becomes an
oscillatory damped mode( for $T>3\times 10^4$ $K$ , in Fig 2b ) . The above
results also hold for larger wave numbers, as it can be seen in Figs. 2c and
2d , where the rates obtained from IP dispersion equation and from the fifth
order polynomial (20), respectively, have been plotted as functions of $k$
for a temperature $T=3.5\times 10^4$ $K$. The fourth order polynomial gives
an oscillatory mode unstable for $k<2.\times 10^{-4}($ critical value ) and
two real modes, the slowest one being unstable. The fifth order polynomial
also gives the above modes but increased by a factor 2, and an additional
very fast damped real mode. The physical reason for the damping of the
oscillatory mode at large enough values of $k$ rests in the second viscosity
originated by the chemical reaction ( Landau \& Lifshitz 1987 ).

The effect of thermal conduction is twofold: it increases the damping and
quenches the instabilities for any fluctuation with wave number $k$ greater
than a critical value $k_c$.

\subsection{Photoionized Hydrogen Plasma with Metallicity Z}

In this section the results of section 3 will be applied to the photoionized
hydrogen model studied by Corbelli and Ferrara (\ CF\ ), i.e. an optically
thin hydrogen plasma with metallicity $Z$ heated and ionized by a background
radiation field of mean photon energy $E$ and ionization rate $\zeta .$ The
net rate function $X(\rho ,T,\xi )$ and the net cooling rate per unit mass $%
\mathcal{L}(\rho ,T,\xi )$ are respectively given by 
\begin{equation}
X(\rho ,T,\xi )=N_0\rho [\xi ^2\alpha -(1-\xi )\xi \gamma _c]-(1-\xi
)(1+\phi )\zeta ,  \label{49}
\end{equation}
\begin{equation}
\mathcal{L}(\rho ,T,\xi )=N_0^2\rho [(1-\xi )Z\Lambda _{HZ}+\xi Z\Lambda
_{eZ}+(1-\xi )\xi \Lambda _{eH}+\xi ^2\Lambda _{eH^{+}}]-N_0(1-\xi )\zeta
[E_h+(1+\phi )\chi ].  \label{50}
\end{equation}
Except for the ionization rate indicated above by $\zeta $, the remaining
notation is like the one used by CF, as well as the corresponding
expressions for: the number of secondary electrons $\phi $ and heat released
per photoionization $E_h$ ( Shull \& Van Steenberg 1985 ), the cooling
efficiencies by collisions neutral hydrogen-ions and metal atoms $\Lambda
_{HZ}$ (Launay \& Roueff 1977, Dalgarno and McCray 1972 ), electrons-ions
and metal atoms $\Lambda _{eZ}$ ( Dalgarno and McCray 1972 ), Ly$\alpha $
emission by neutral hydrogen $\Lambda _{eH}$ (\ Spitzer 1978 ) and hydrogen
recombination $\Lambda _{eH^{+}},$ on the spot approximation (Seaton 1959 ).

By solving numerically the fifth order dispersion equation (20), one obtains
the following results: for a radiation field with a photon energy $E=15$ $eV$%
, a plasma with $Z=1$ becomes stable in the whole range of temperature ( $%
T\leq 8.26\times 10^2K$ ) where the thermochemical equilibrium may exist,
contrary to the CF result where there is instability in the range $%
33<T<703K. $ Figures 3a and 3b show the resulting rates, for a disturbance
with $k=1$ ( in units of $10^{-18}cm^{-1}$ ), by solving the fourth [Eq.
(2.7) of CF\ ] and fifth [Eq. (20) of Sec. 2 above ] order polynomials,
respectively. In the first case, there is a damped mode ( the fastest one )
with two bifurcation points and an unstable oscillating mode in the above
range of temperature. The dispersion equation (20) leaves an additional non
oscillating damped mode with the highest rate, unfolds the damped
oscillating mode ( of Fig 3a ) in two damped real modes, and stabilizes the
oscillating unstable mode (of Fig.3a ) on the whole range of temperature.
The rates corresponding to the above modes also are shown in Figs. 3c and
3d, as functions of the wave number $k$, for a temperature of $T=60K$ (
temperature at which $\xi _{*T}$ becomes a maximum ). The slowest
oscillating mode becomes unstable for $k<14.5$ in Fig 3c; instead, it
remains stable in Fig.3d with a rate being an increasing function of $k$.
The maximum damping occurs for acoustic oscillations with $k>10^3$. This
effect is expected to occur as a consequence of the second viscosity due to
the chemical reaction (Landau and Lifshitz 1987). Additionally, the new
damped non- oscillating mode shows dispersion at high values of $k$ ( $%
>3\times 10^2$). Similar behavior is found at other values of temperature.

Fig 4a (fourth order polynomial ) and 4b (fifth order polynomial ) show the
rates as function of temperature for disturbances with $k=1$, in a plasma
with $Z=1,$ and photon energy $E=10^2$ $eV$. With the exception of appearing
the fifth non oscillating damped mode ( with a rate value between the
corresponding rates values of the two modes given by the fourth order
polynomial ), no qualitative changes appear. Only slight changes are
observed in the values of temperature at which bifurcation occurs for both
modes, the fastest stable and the slowest unstable one. In Figs 4c (fourth
order polynomial ) and 4d (fifth order polynomial ) the corresponding rates
are plotted as functions of the wave number $k$, for a temperature $T=200K.$
The rates of the damped mode and the corresponding value of $k$ where
bifurcation occurs are shifted to high values while the unstable oscillating
mode ( Fig 3c ) is stabilized for wave numbers $k>4.8$ ( Fig.4d\ ).

Figs. 5a, 5b are as Figs. 3a, 3b but for $Z=0.5$; and Figs. 5c, 5d are as
Figs. 3c and 3d but for a temperature $T=500K$ $.$ There, in the whole range
of $k$ under consideration the fourth order polynomial gives two oscillating
modes, the fastest one being stable and the slowest one unstable for $k<4.8$
( Fig. 5c ). Instead, the fifth order polynomial unfolds the above damped
oscillating mode in two non oscillating modes and stabilizes the unstable (
slowest ) mode. Additionally, the new non- oscillating damped mode also
shows dispersion at high wave values of $k$ ( $>10^2$ ), and the acoustic
mode is also strongly damped by second viscosity for values of $k>$ $2\times
10^2$ ( Fig.5d ). Similar results hold for other values of the temperature.

For low metallicity plasmas ( say $Z=10^{-3}$ ), the fifth order polynomial
gives three non oscillating and one oscillating mode, or two oscillating and
one real mode. In particular, for $E=100eV,$ and in the range temperature
where the plasma is unstable, i.e. $72<T<4.34\times 10^3K$, the fourth order
polynomial gives a stable and an unstable real mode ( the slowest one ), and
a damped acoustic mode. Instead, the fifth order polynomial gives an
unstable real mode ( also the slowest one ) and two damped acoustic modes.
This unstable thermal mode is slightly faster than the corresponding one
resulting from solving the fourth order polynomial. On the other hand, for $%
Z=0$, the plasma also becomes thermochemically stable.

Specifically, in clouds located outside the optical disk of galaxies or in
the galactic halo the probable values for the energy and metallicity, quoted
out by CF are: $E=20$ $eV$ , $Z=1$ and $T\approx 800$ $K.$ For these values
the corresponding rates have been plotted in Figures 6a-6b ( as functions of 
$T$ ) and 6c-6d ( as functions of $k$ ). The range of temperature for which
oscillatory instability may set in according to the CF analysis is rather
wide $43<T<3.21\times 10^3$ $K,$ ( Fig. 6a ) , instead, the present analysis
gives a considerable reduction of it, i.e. the plasma is unstable only in
the range of temperature $49<T<8.46\times 10^2$ $K.$ Additionally, while the
fourth order dispersion equation gives oscillatory instability for any value
of the wave number and maximum growing rates $\approx 10^{-5}$ $yr^{-1}$ ,
for disturbances with $k>10^{-17}cm^{-1}$( Fig 6c\ ), from the dispersion
relation (20) one obtains that there is instability only for fluctuation
with $k<k_c$ ( Fig. 6d ), where the critical value $k_c$ is weakly dependent
on temperature ( $10^{-18}<k_c<9.6\times 10^{-18}$ $cm^{-1}$ ), in the above
range of temperature where the plasma is unstable ). The maximum growing
rate is strongly dependent on $T$, and the upper value $n_{\max }=4.8\times $
$10^{-7}yr^{-1}$is reached at temperature $T\approx 500\;K$ for fluctuations
with wave number $k=6.0\times 10^{-18}$ $cm^{-1}$ (see Fig.6d , dashed line
).

>From the above , one may conclude that the effect of taking into account the
change of the ionization equilibrium position due to the variation of
density and temperature, is to stabilize the oscillating modes, at least for
high enough values of the wave number, as it should be expected from
Landau's results on second viscosity. However, the quantitative stabilizing
effect depends on the particular values of photon energy , metallicity and
plasma temperature. For instance, for the values used in Figures 3c-3d and
5c-5d, such an effect is able to stabilize the unstable acoustic mode in the
whole range of $k$ under consideration. But, for those used in Fig4c-4d the
effect is only capable of stabilizing acoustic oscillations with $k>4.8$.

\section{Summary and Conclusions}

In summary, if one takes into account the effects of a finite chemical
relaxation time in optically thin reacting plasmas a fifth order dispersion
equation is obtained, instead of the fourth order polynomial obtained when
such effects are neglected. The quantitative effects on the previous four
modes depend on the particular reacting plasma under consideration, but
generally, it tends to further stabilize the plasma and to increase the
damping of stable modes because of the second viscosity generated by the
chemical reaction.

In the collisionally ionized pure hydrogen plasma model previously studied
by IP the resulting effects are not too strong. In addition to the
appearance of the new damped real mode, only an increase in the value of
both the damping and growing rates occur. However, in the photoionized
plasma studied by CF, the effects may result more severe, depending on the
exact value of the parameters under consideration. In particular, the range
of temperature where the plasma is unstable can be strongly reduced, and the
oscillatory instability can be quenched for $k$ large enough. So for
instance, for values likely occurring in the galactic halo, the present
analysis leaves a very restrict values of $k$ for developing instability.
See the end of the previous Section.

\section*{Acknowledgments}

The author thanks the Referee Dr. A. Ferrara for his helpful criticisms.
This work has been partially supported by the Consejo de Desarrollo
Cient\'{i}fico, Human\'{i}stico y Tecnol\'{o}gico (CDCHT) de la Universidad
de los Andes through Project C-750-95-B05.

\newpage
FIGURE CAPTIONS

Fig. 1. The equilibrium pressure as a function of particle number density
for a collisionally ionized pure hydrogen plasma. The temperatures of
marginal states are indicated by the labels $T_1...T_4$, see text.

Fig. 2. The resulting rates from solving the fourth order polynomial of IP
as functions of temperature for a perturbation with wave number $k=10^{-4}$
(a), and as functions of $k$ for a temperature $T=3.5\times 10^4$ $K$ (c).
The resulting rates from the fifth order polynomial (20) as functions of $T$
for $k=10^{-4}$\ (b), and as functions of $k$ for $T=3.5\times 10^4$ $K$
(d). For a heating rate value of the order of the galactic heating rate, $k$
is in units of $10^{-18}$ $cm^{-1}$. Positive rates of complex modes are
indicated by $C^{+}$ and those corresponding to real roots by $R^{+}$.

Fig.3. The rates resulting from solving the fourth order (a) an fifth order
(b) polynomials ( see text ) as functions of temperature for, $N_0\rho =1$ $%
cm^{-3}$, a photon energy $E=15$ $eV,$ metallicity $Z=1$, and $k=1$. The
wave number $k$ is in units of $10^{-18}$ $cm^{-1}$. The values of the two
negative intermediate real rates are indistinguishable at the scale of Fig.
(b) . Figures (c) and (d) respectively correspond to the rates on Figures
(a) and (b) but as functions of $k$, and for $T=60$ $K$. As in Figure (b)
the two damped intermediate real roots are also indistinguishable at the
scale of Fig. (d). The range where the oscillating mode is unstable is
indicated by $C^{+}$.

Fig. 4. Figures (a) and (b) are as Figures 3a and 3b, respectively, but for $%
E=100$ $eV$ . Figures (c) and (d) are as Figures 3c and 3d, respectively,
but for $E=100$ $eV$ and $T=200$ $K$. The range where the real mode is
unstable is indicated by $R^{+}$.

Fig. 5. Figures (a) and (b) are as Figures 3a and 3b, respectively, but for $%
Z=0.5$ . Figures (c) and (d) are as Figures 3c and 3d, respectively, but for 
$Z=0.5$ and $T=500$ $K$.

Fig. 6. Figures (a) and (b) are as Figures 3a and 3b, respectively, but for $%
E=20$ $eV$, $Z=1$ . Figures (c) and (d) are as Figures 3c and 3d,
respectively, but for $E=20$ $eV$, $Z=1$ and $T=800$ $K$. On Fig (d) the
dashed curve corresponds to a temperature $T=500$ $K$, value at which the
maximum growing rate is reached.


\begin{thebibliography}{99}
\bibitem{}  Baron, E., Carswell, R. F., Hogan, C. J., \& Weymann, R. J.
1989, ApJ, 337, 609

\bibitem{}  Corbelli, E., \& Ferrara, A. 1995, ApJ, 447, 720

\bibitem{}  Cowie, L. L., McKee, C. F., \& Ostriker, J. P. 1981, ApJ, 247,
908

\bibitem{}  Charlton, J. C., Salpeter, E. E., Hogan, C. J. 1993, 402, 493

\bibitem{}  Dalgarno, A., \& McCray, R. A. 1972, ARA\&A, 10, 375

\bibitem{}  Defouw, R. J. 1970, ApJ, 161, 55

\bibitem{}  Fall, S. M., \& Rees, M. 1985, ApJ 298, 18

\bibitem{}  Field, G. B. 1965, ApJ, 142, 531

\bibitem{}  Ferrara, A., \& Field, G. B. 1994, ApJ, 423, 665

\bibitem{}  Flannery, B. P., \&Press, W. H. 1979, ApJ, 231, 688

\bibitem{}  Glassgold and Lager 1976, A. E., \& Lager, W. D. 1976, 204, 403

\bibitem{}  Goldsmith, G. B. 1970, ApJ, 161, 41


\bibitem{}  Goldsmith, G. W., Habing , H. J., \& Field, G. B.1969, ApJ, 158,
173

\bibitem{}  Goncalves, D. R., Jatenco-Pereira, V., \& Opher, R, 1993, ApJ,
414,57

\bibitem{}  Gurevich, L. E.,\ \& Chernin, A. D. 1975, AZh, 19,1

\bibitem{}  Hummer, D. G. 1963, MNRAS, 125, 461

\bibitem{}  Hummer, D. G., \& Seaton, M. J. 1963, MNRAS, 125, 437

\bibitem{}  Hunter, J. H. 1970, ApJ, 161, 451

\bibitem{}  Iba\~{n}ez , S. M. H.1981, MNRAS, 196, 13

\bibitem{}  Iba\~{n}ez, S. M. H., \& Parravano, A. 1983, ApJ, 275, 181

\bibitem{}  Iba\~{n}ez, S. M. H., \& Mendoza, B. C. A. 1990, Ap\&SS, 164, 193

\bibitem{}  Krolik, J. H., McKee, C. F., \& Tarter, C, B. 1981, ApJ, 249, 422

\bibitem{}  Krolik, J. H. 1988, ApJ, 325, 148

\bibitem{}  Kulkarni, V. S., \&Fall, S. M. 1993, ApJ, 413, L63

\bibitem{}  Kwan, J. Y., \& Krolik, J. H. 1981, ApJ, 250, 468

\bibitem{}  Landau, L. D., \& Lifshitz, E. M. 1987, Fluid Mechanics (London:
Pergamon Press ), 308

\bibitem{}  Launay , J. M., \& Roueff, E. 1977, A\&A,56, 289

\bibitem{}  Mathews, W. G. 1986, ApJ, 305, 187Seaton, M. J. 1959, 119, 81

\bibitem{}  Mathews, W. G., \& Doane, J. S. 1990, ApJ, 352, 423

\bibitem{}  McKee, C. F., \& Ostriker, J. P. 1997, ApJ, 218, 148

\bibitem{}  Murakami, I., \& Ikeuchi, S. 1990, PASJ, 42, L11

\bibitem{}  Murray, S. D., \& Lin, D. N. C. 1990a, ApJ, 357, 105

\bibitem{}  Murray, S. D., \& Lin, D. N. C. 1990b, ApJ, 363, 50

\bibitem{}  Parker, E. 1953, ApJ, 117, 431

\bibitem{}  Potach, S. R., Wesselius, P. R., \& Duinen, R. J. 1979, A\&A,
74, L15

\bibitem{}  Priest, E. 1982, Solar Magnetohydrodynamics ( Dordrecht: reidel
), 235, 278, 382

\bibitem{}  Reale, F., Serio, S., \& Peres, G. 1994, ApJ, 433,811

\bibitem{}  Shull, J. M., \& Van Steenberg, M. E. 1985, ApJ, 298, 268

\bibitem{}  Spitzer, L. 1962, Physics of Fully Ionized Gases ( New York:
Wiley )

\bibitem{}  Spitzer, L. 1978, Physical processes in the Interstellar Medium
( new York: Wiley )

\bibitem{}  Sunyaev, R. A., \& Zel'dovich, Ya. B. 1972, A\&A, 20, 200

\bibitem{}  Verschuur, G. L., \& Magnan, 1994, AJ, 107, 287

\bibitem{}  Vicenti, W. G., \& Kruger, Ch. H. 1975, Introduction to Physical
Gas Dynamics (\ New York:\ Wiley ), 236

\bibitem{}  Yoneyama, T.1973, PASJ, 25, 349

\bibitem{}  Zel'dovich, Ya. B., \& Novikov, I. D. 1983, The Structure and
Evolution of the Universe ( Chicago: Univ. of Chicago Press ), 344

\bibitem{}  Zel'dovich, Ya. B., \& Rayzer, Yu. R. 1967, Physics of Shock
Waves and High-Temperature Hydrodynamic Phenomena ( New York: Academic Press
) , 349
\end{thebibliography}
\end{document}